\documentclass[usenatbib,usegraphicx]{mn2e}

\title[Pulsar force-free magnetosphere close to the magnetic axis]
{Axisymmetric force-free magnetosphere of a pulsar. I. The structure close to the magnetic axis}

\author[S. A. Petrova]{S. A. Petrova
\thanks{E-mail: petrova@ri.kharkov.ua}\\
Institute of Radio Astronomy, NAS of Ukraine, Chervonopraporna
Str., 4, Kharkov 61002, Ukraine}

\begin{document}

\date{Received\dots}

\pagerange{\pageref{firstpage}--\pageref{lastpage}} \pubyear{2012}

\maketitle

\protect\label{firstpage}

\begin{abstract}
The stationary axisymmetric force-free magnetosphere of a pulsar is studied analytically. The pulsar equation is solved in the region close to the magnetic axis. Proceeding from linearization of the current function in the axial region, we find the axial magnetic flux function valid at any altitude above the neutron star. This function is used as a starting approximation to develop series for the non-linear pulsar equation in the polar region. Taking into account the quasi-monopolar character of the pulsar magnetic flux at infinity, we obtain unique asymptotic series for the flux and current functions. At infinity, both functions are close but not equivalent to those known for the case of a force-free monopole. The flux function at the top of the polar gap is found to differ from the dipolar one at the neutron star surface. With our results, the transverse current sheet closing the pulsar circuit at the neutron star surface is consistently incorporated into the global magnetospheric structure, the backward particle flow at small polar angles can be excluded and the stationary cascade scenario looks admissible. The present paper is the first step toward complete analytic description of the pulsar force-free magnetosphere allowing for the plasma-producing gaps and pulsar current circuit closure.
\end{abstract}


\section{Introduction}

Pulsar magnetosphere contains the plasma \citep{gj69}, which is sufficiently abundant to affect the magnetic field structure. This necessitates a self-consistent consideration of fields and currents in pulsars. The basic model of the pulsar magnetosphere is that of a rotating axisymmetric force-free dipole, where the magnetic and rotational axes are aligned, the electromagnetic forces are balanced and the particle inertia is ignored. Then the poloidal current and magnetic flux are related by the well-known pulsar equation \citep{m73,sw73,o74}. Being formulated almost 40 years ago, it still lacks a proper analytic solution for the dipolar case.

An exact solution is known only for the force-free monopole \citep{m73}, including the versions of a split and offset monopole \citep{m91,p12}, which are believed to replicate the behaviour of the force-free dipole at large distances from the origin. In the case of a monopole, the problem is strongly simplified, since the poloidal component of the force-free magnetic field appears to be the same as in the original vacuum configurations. In the case of a dipole, the flux and current functions entering the pulsar equation are both unknown, and the problem is to guess such a current distribution that makes the self-consistent flux function obey the physically meaningful boundary conditions. The early attempts of solving the pulsar equation for the current functions of a special form yielded the flux functions which were valid only inside the light cylinder and could not be smoothly continued to infinity \citep*{m73b,bgi83}. Later on a more sophisticated current function providing a smooth passage of the magnetic field lines across the light cylinder was found by means of numerical methods \citep*{ckf99}. The necessity to revise the set of commonly used boundary conditions was pointed out in \citet{p12}.

The pioneering work of \citet{ckf99} inspired extensive numerical studies of the pulsar equation. The numerical solution for the stationary axisymmetric force-free magnetosphere was confirmed by the time-dependent simulations \citep{k06,mck06,s06} and generalised in a number of aspects. In particular, the non-axisymmetric 3D magnetosphere was considered \citep*{s06,kc09,bs10,kck12}, the differential rotation was taken into account \citep{c05,t06,t07}, the non-ideal MHD case was addressed \citep*{kkhc12,lst12}. However, the set of boundary conditions used in all these simulations did not allow for the presence of the plasma producing gaps. Therefore the numerical treatment of the pulsar equation was not self-consistent.

In our previous work \citep{p12}, we have suggested an empirical model of the pulsar axisymmetric force-free magnetosphere, which for the first time includes the polar, outer and slot gaps into the global magnetospheric structure. The basic prescriptions of the model are as follows. Although the polar gap height can be regarded as infinitesimal, the magnetic field structure at its top differs from the original dipolar one at the neutron star surface. As the outer gap is entirely located within the light cylinder, beyond the light cylinder the null line should not intersect the magnetic field lines. The polar and outer gaps control different bundles of the open field lines and produce the direct and return currents, respectively. Therefore beyond the light cylinder the null line coincides with the critical field line, both going parallel to the magnetic equator at a certain altitude above it. The compensation current necessary to close the pulsar circuit originates in the slot gap located along the critical field line inside the light cylinder.

Our qualitative model needs to be quantified. With the present paper, we begin a systematic analytic description of the pulsar magnetosphere within the framework of this model. The present study is focused on the region close to the magnetic axis. The absence of the poloidal current along the magnetic axis provides an essential simplification: in the nearest vicinity of the axis, the generally non-linear current function and the pulsar equation on the whole can be linearised. In Sect.~2, we solve the simplified pulsar equation in the axial region and find the magnetic flux function at any altitude $z$ above the neutron star. In Sect.~3, the complete non-linear pulsar equation is considered. We search for the solution in terms of series in $\rho^2/z^2$  (where $\rho$ is the separation from the axis) and use the axial flux function as a starting approximation. Taking into account the quasi-monopolar character of the flux function at infinity, we find a unique current function and the corresponding unique asymptotic form of the flux function at large altitudes above the neutron star. With the current function in hand, we also obtain the flux function at the top of the polar gap and demonstrate its distinction from that of a dipole. Our results and their implications are discussed in Sect.~4 and briefly summarised in Sect.~5.

\section{Simplified pulsar equation in the axial region} 
\protect\label{s2}

Our consideration concerns the model of a stationary axisymmetric force-free dipole. It is convenient to choose the cylindrical coordinate system $(\rho,\phi,z)$ with the axis along the pulsar axis. Then the pulsar equation reads \citep{m73,sw73,o74} 
\begin{equation}
\left (1-\rho^2\right )\left[\frac{\partial^2f}{\partial\rho^2}+\frac{1}{\rho}\frac{\partial f}{\partial\rho}+\frac{\partial^2f}{\partial z^2}\right]-\frac{2}{\rho}\frac{\partial f}{\partial\rho}=-A\frac{\mathrm{d}A}{\mathrm{d}f},
\label{eq1}
\end{equation}
where the dimensionless functions $f(\rho,z)$ and $A(f)$ are proportional respectively to the magnetic flux and poloidal electric current through the circle of a radius $\rho$ centered at the magnetic axis at an altitude $z$. As we are interested in the region close to the magnetic axis, the relevant boundary conditions are as follows. Along this axis, the flux $f$ and current $A$ are both zero. At infinity, the field lines become radial, $f=f(\rho/z)$, and resemble those of a magnetic monopole. According to the current models \citep*[e.g.][]{s02,s12}, the pulsar current circuit closes through the neutron star. Then out of the circuit, well below the stellar surface, the poloidal current is zero, the magnetic field is dipolar, $f=\rho^2/(\rho^2+z^2)^{3/2}$, and satisfies the multipolar equation
\begin{equation}
\frac{\partial^2f}{\partial\rho^2}-\frac{1}{\rho}\frac{\partial f}{\partial\rho}+\frac{\partial^2f}{\partial z^2}=0.
\label{eq2}
\end{equation}
In the region of interest, $\rho\ll 1$, we have $f\approx\rho^2/z^3$. Just above the top of the transverse current sheet, the current is aligned with the poloidal magnetic field and remains constant along a fixed field line. Note that the jump of the poloidal current from zero to the value dictated by the global force-free configuration, which occurs at the neutron star surface bearing the transverse current, is compatible with the continuity equation, $\nabla\cdot \bmath{J}=0$.

The presence of the infinitesimally narrow polar gap just above the transverse current sheet nothing affects the formulation of the force-free problem. Indeed, both unknowns, $A$ and $f$, are not changed in the course of the particle acceleration, and the role of the gap is only to supply a sufficient amount of plasma necessary to sustain the force-free configuration. Therefore in our consideration the top of the polar gap and, correspondingly, the bottom of the force-free region can be assumed to coincide with the top of the transverse current sheet at the stellar surface. In the present paper, we also neglect the potential drop in the polar gap and the resultant differential rotation of the magnetosphere.

The current function on the right-hand side of Eq.~(\ref{eq1}) is unknown. It is believed to be non-linear, in order to provide the current circuit closure, and close enough to that of a force-free monopole, $A\mathrm{d}A/\mathrm{d}f=2f(1-f/f_0)(2-f/f_0)$, where $f_0$ is a constant \citep[see, e.g.,][]{c05}. Note that the solution of the pulsar equation with an assumption of a linear current function \citep{bgi83} appeared non-physical, since it could not continuously cross the light cylinder. Indeed, the solution of a non-linear problem in general differs essentially from that of a linear one. At the same time, in the nearest vicinity of the magnetic axis, the non-linear current function can be expanded into the Taylor series in $f$. Keeping in mind its similarity to the current function of a force-free monopole, one can expect that the linear term of the expansion is non-zero. Thus, close to the magnetic axis, the current function can be linearised.

With the above assumptions, the pulsar equation is substantially simplified and takes the form
\begin{equation}
\frac{\partial^2f}{\partial\rho^2}-\frac{1}{\rho}\frac{\partial f}{\partial\rho}+\frac{\partial^2f}{\partial z^2}=-\alpha f\chi(z-z_0),
\label{eq3}
\end{equation}
where $\alpha$ is an arbitrary constant, $\chi(z-z_0)$ is the step function, $z_0$ stands for the polar gap altitude, and the terms $\sim\rho^2$ are ignored. Strictly speaking, for these terms to be ignored the step function $\chi(z-z_0)$ should actually be somewhat smeared at a scale $h\gg\rho$. Thus, assuming a narrow transverse current sheet, $h\ll z_0$, we are restricted to consider only the nearest vicinity of the magnetic axis, $\rho\ll h$, which we call the axial region.

Eq.~(\ref{eq3}) allows separation of variables, in which case the solution can be presented as
\begin{equation}
f=\int_0^\infty C(\lambda)v(\rho)u(z)\mathrm{d}\lambda,
\label{eq4}
\end{equation}
where the functions $v(\rho)$ and $u(z)$ obey the equations
\begin{equation}
\frac{\mathrm{d}v}{\mathrm{d}\rho^2}-\frac{1}{\rho}\frac{\mathrm{d}v}{\mathrm{d}\rho}+\lambda^2v=0,
\label{eq5}
\end{equation}
\begin{equation}
\frac{\mathrm{d}u}{\mathrm{d}z^2}-\lambda^2u=-\alpha u\chi(z-z_0),
\label{eq6}
\end{equation}
$\lambda$ is independent of $\rho$ and $z$, and $C(\lambda)$ is an arbitrary function.
Since at the magnetic axis $f=0$, we have $v(\rho)=\rho J_1(\lambda\rho)$. In order to provide the dipolar character of the magnetic field at $z<z_0$, $f=\rho^2/(\rho^2+z^2)^{3/2}$, it is necessary to choose $u(z)=\exp(-\lambda z)$ and $C(\lambda)=\lambda$. In the Appendix we solve Eq.~(\ref{eq6}) for the smoothed step function and show that well above the top of the polar gap $u(z)=\exp(-z\sqrt{\lambda^2-\alpha})$. Then the flux function reads
\begin{equation}
f=\frac{\rho^2}{2}\int_0^\infty\sqrt{\mu^2+\alpha}\,\mathrm{e}^{-\mu z}\mu\mathrm{d}\mu,
\label{eq7}
\end{equation}
where $\mu^2=\lambda^2-\alpha$ and it is taken that $J_1(\lambda\rho)\approx\lambda\rho/2$.

Eq.~(\ref{eq7}) can be calculated using the integral \citep{gr80}
\[
\int_0^\infty(x^2+u^2)^{\nu-1}\mathrm{e}^{-px}\mathrm{d}x=\frac{\sqrt{\pi}}{2}\left(\frac{2u}{p}\right)^{\nu-1/2}\Gamma(\nu)
\]
\begin{equation}
\times\left[\bmath{H}_{\nu-1/2}(up)-Y_{\nu-1/2}(up)\right],
\label{eq8} 
\end{equation}
where $\Gamma(\nu)$ is the Euler gamma-function, $\bmath{H}_{\nu-1/2}(up)$ the Struve function, and $Y_{\nu-1/2}(up)$ the Bessel function of the second kind. Differentiating both sides of Eq.~(\ref{eq8}) with respect to $p$ and making use of the recurrent relations for the Struve and Bessel functions yield
\[f=\rho^2\Phi(z),\]
with
\begin{equation}
\Phi=-\frac{\alpha^{3/2}}{6}+\frac{\pi\alpha}{4z}\left[\bmath{H}_2\left(\sqrt{\alpha}z\right)-Y_2\left(\sqrt{\alpha}z\right)\right].
\label{eq9}
\end{equation}
Thus, we have found the axial flux function of the rotating axisymmetric force-free dipole. In order to confirm the validity of our technique of simplifying and solving the pulsar equation, one can analogously apply it to the force-free monopole, in which case the exact solution is known. The monopolar condition at the origin, $f=1-z/\sqrt{z^2+\rho^2}$, implies that $C(\lambda)=1$. Then, instead of Eq.~(\ref{eq7}), we have
\[
f=\frac{\rho^2}{2}\int_0^\infty\mathrm{e}^{-\mu z}\mu\mathrm{d}\mu,
\]
which is exactly reduced to $f=\rho^2/2z^2$, just as could be expected.

Taking into account the approximations at $z\to 0$,
\[
\bmath{H}_\nu(\xi)\sim\frac{(\xi/2)^{\nu+1}}{\Gamma(3/2)\Gamma(\nu+3/2)},\quad Y_\nu(\xi)\sim-\frac{1}{\pi}\Gamma(\nu)(\xi/2)^{-\nu},
\]
one can see that at $z\ll 1$ the magnetic field structure is roughly dipolar, $f\sim\rho^2/z^3$ (recall that we consider the region very close to the magnetic axis, $\rho/z\ll 1$). With the asymptotic relation
\[
\bmath{H}_\nu(\xi)-Y_\nu(\xi)=\frac{1}{\pi}\sum_{k=0}^{m-1}\frac{\Gamma(k+1/2)}{\Gamma(\nu+1/2-k)(\xi/2)^{2k-\nu+1}}\]
\begin{equation}
+O(\vert z\vert)^{\nu-2m-1} 
\label{eq10}
\end{equation}
we find that at large altitudes, $z\gg 1$, the flux function bears the monopolar character, $f\sim\sqrt{\alpha}\rho^2/2z^2$. This is just what is expected for the force-free dipolar field and attributed to the action of the current, which flows through the polar gap and further along the open magnetic lines. Our consideration shows that, at least in the axial region, the flux function of a force-free dipole asymptotically coincides with $f_{\mathrm{mon}}=\sqrt{\alpha}(1-z/\sqrt{\rho^2+z^2})$.

\begin{figure*}
\includegraphics[width=190mm]{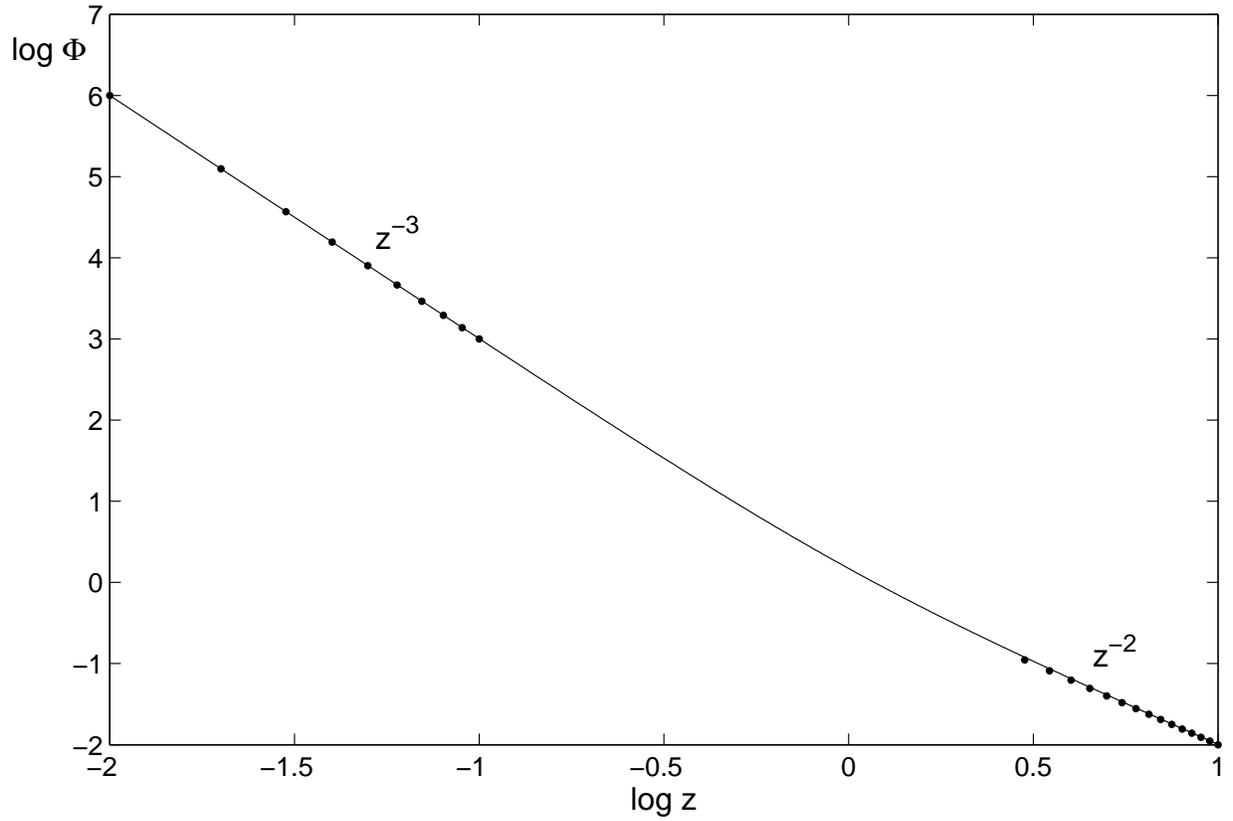}
\caption{The axial magnetic flux as function of the altitude above the neutron star based on Eq.~(\ref{eq9}). The approximations at $z\to 0$ and $z\to\infty$ are shown by dots.}
\label{f1}
\end{figure*}

\begin{figure*}
\includegraphics[width=190mm]{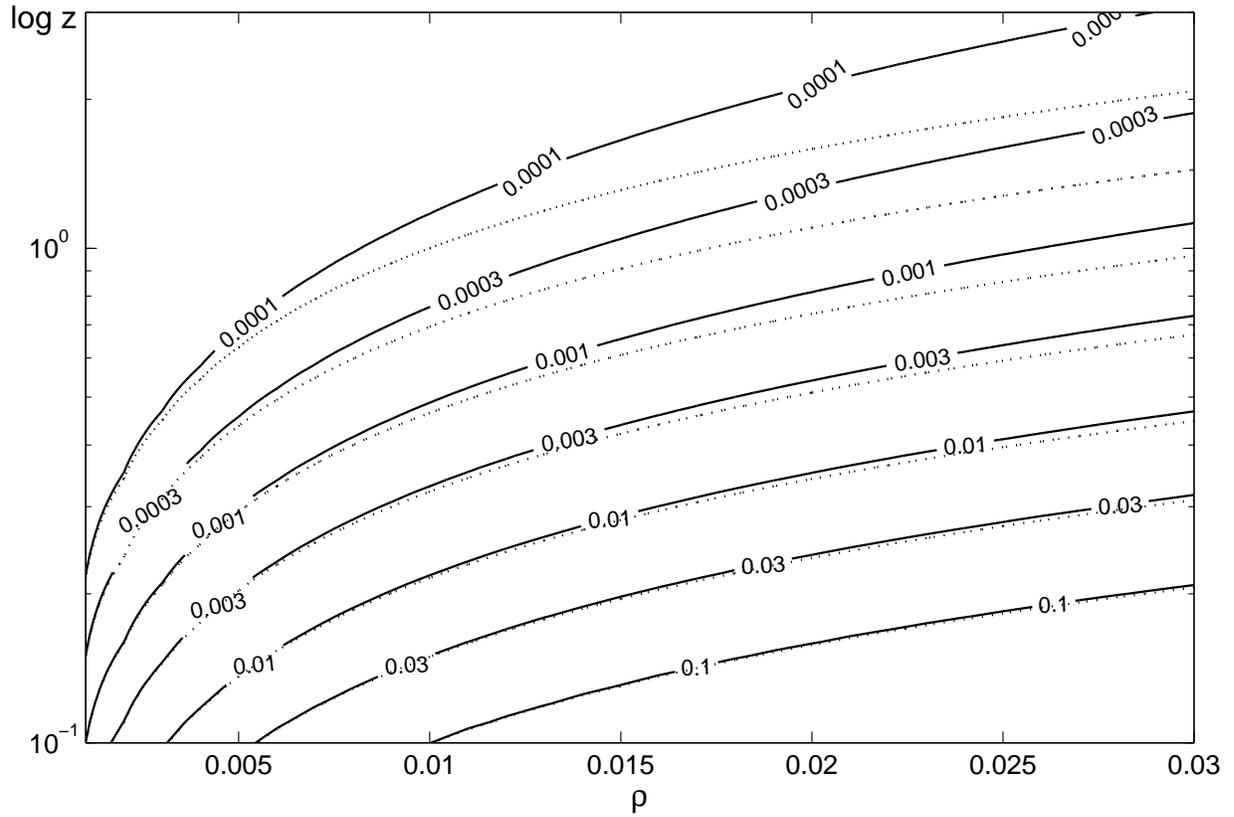}
\caption{Field lines of the force-free magnetic field in the axial region of a pulsar (solid lines). Dotted lines correspond to the dipolar case.}
\label{f2}
\protect\label{lastpage}
\end{figure*}

The function $\Phi(z)$ along with its asymptotes is plotted in Fig.~\ref{f1}. Figure \ref{f2} shows the magnetic field lines in the axial region for the cases of a force-free and vacuum dipole. One can see that the force-free field lines tend to go closer to the axis. This seems to be in line with the general fact that the force-free dipole has an additional bundle of open field lines as compared to the vacuum dipole \citep[e.g.,][]{m73b,ckf99,c05}.

\section{Complete pulsar equation at small polar angles} 
\protect\label{s3}

In the previous section, we have obtained the axial flux function for the rotating axisymmetric force-free dipole. With this result in hand, we shall treat the complete pulsar equation (\ref{eq1}) at small polar angles, $\theta=\mathrm{atan}(\rho/z)\ll 1$. The solution can be searched for in the form
\begin{equation}
f=\sum_{k=1}^\infty b_k(z)\rho^{2k}/z^{2k},
\label{eq11}
\end{equation}
where $b_1(z)=z^2\Phi(z)$. Substituting Eq.~(\ref{eq11}) into Eq.~(\ref{eq1}) leads to the recurrent relation
\[
4k(k+1)\frac{b_{k+1}}{z^{2k+2}}-4k^2\frac{b_k}{z^{2k}}+\frac{\mathrm{d}^2}{\mathrm{d}z^2}\left(\frac{b_k}{z^{2k}}\right)-\frac{\mathrm{d}^2}{\mathrm{d}z^2}\left(\frac{b_{k-1}}{z^{2k-2}}\right)\]
\begin{equation}=-AA^\prime_{2k},
\label{eq12}
\end{equation}
where
\[
AA^\prime_{2k}=\frac{1}{z^{2k}}
\]
\begin{equation}
\times\left(\xi_1b_k+\xi_2\sum_{i=1}^{k-1}b_ib_{k-i}+\xi_3\sum_{i=1}^{k-2}\sum_{j=1}^{k-i-1}b_ib_jb_{k-i-j}\dots\right)
\label{eq13}
\end{equation}
and $\xi_i$ are constants. The latter relation results from expanding $AA^\prime$ into the Taylor series in $f$ and using Eq.~(\ref{eq11}).

If the current function were given and, correspondingly, the set of $\xi_i$ were known, one would successively obtain the functions $b_k$ from Eq.~(\ref{eq12}). However, in our case the flux and current functions are both unknown and should be found self-consistently. For this purpose we additionally impose the boundary condition at infinity. Namely, at infinity the magnetic field lines should become radial, i.e. the flux function should only be the function of the polar angle, $f=f(\rho/z)$, and, correspondingly, for any $k$
\begin{equation}
\lim_{z\to\infty}b_k(z)=\mathrm{Const}_k.
\label{eq14}
\end{equation}
Applying this condition to Eq.~(\ref{eq12}) at each step of the iteration procedure and using the asymptotic representation (\ref{eq10}), one can obtain a unique value of $\xi_k$ and the corresponding function $b_{k+1}(z)$, which is also unique. In practice, it is sufficient to proceed directly from the asymptotic form of $b_1(z)$,
\begin{equation}
b_1(z)\approx 1+\frac{3}{4z^2}-\frac{15}{16z^4}+\frac{315}{64z^6}-\dots,
\label{eq15}
\end{equation}
retaining $k$ terms in order to find the asymptotic value of $b_{k+1}$. As a result, the current function of an axisymmetric force-free dipole and the asymptotic form of the corresponding flux function at $z\to\infty$ and $\rho/z\ll 1$ read
\[
AA^\prime=4f-3f^2-\frac{7}{2}f^3+\frac{675}{16}f^4+\dots,
\]
\begin{equation}
f=\frac{\rho^2}{z^2}-\frac{3}{4}\frac{\rho^4}{z^4}+\frac{1}{2}\frac{\rho^6}{z^6}+\frac{13}{32}\frac{\rho^8}{z^8}+\frac{9567}{1280}\frac{\rho^{10}}{z^{10}}+\dots .
\label{eq16}
\end{equation}

Taking the axial flux function for the force-free monopole, $f_{\mathrm{mon}}=\rho^2/z^2$ (in which case $\alpha=4$), we have exactly $b_1=1$ and obtain analogously
\[
AA^\prime_\mathrm{mon}=4f-3f^2+\frac{1}{2}f^3,
\]
\begin{equation}
f_\mathrm{mon}=\frac{\rho^2}{z^2}-\frac{3}{4}\frac{\rho^4}{z^4}+\frac{5}{8}\frac{\rho^6}{z^6}-\frac{35}{64}\frac{\rho^8}{z^8}+\frac{63}{128}\frac{\rho^{10}}{z^{10}}+\dots .
\label{eq17}
\end{equation}
One can see that in these two cases the flux and current functions differ only slightly,
\begin{equation}
f-f_\mathrm{mon}\sim-\frac{1}{8}\frac{\rho^6}{z^6},\quad AA^\prime-AA^\prime_\mathrm{mon}\sim -4f^3,
\label{eq18}
\end{equation} 
but far enough from the magnetic axis this distinction may well become substantial.

We did not manage to guess the analytic form of the functions corresponding to the asymptotic series (\ref{eq16}). While trying different transformations, we have only found that $f$ can be presented in a more suitable form,
\begin{equation}
f=\sin^2\theta+\frac{1}{4}\sin^4\theta+\frac{21}{32}\sin^8\theta+\dots, 
\label{eq19}
\end{equation}
where $\theta=\mathrm{atan}(\rho/z)$. Since the term $\sim\sin^6\theta$ is zero, it is reasonable to keep only the first two terms of the series. Then at $\sin\theta=1$ we have $f=1.25$, which is, unexpectedly, in a good correspondence with the value $f=1.23$ obtained by \citet{c05} numerically.

At large radial distances, $r=\sqrt{z^2+\rho^2}\to\infty$, the asymptotic form of the pulsar equation (\ref{eq1}) reads \citep[e.g.,][]{ckf99}
\begin{equation}
t^2(1+t^2)\frac{\mathrm{d}f}{\mathrm{d}t^2}+(2t^3+t)\frac{\mathrm{d}f}{\mathrm{d}t}=AA^\prime, 
\label{eq20}
\end{equation}
where $t=\rho/z$.  Integration of this equation yields
\begin{equation}
t\sqrt{1+t^2}\frac{\mathrm{d}f}{\mathrm{d}t}=A.
\label{eq21}
\end{equation}
One can check that the functions $AA^\prime$ and $f$ given by Eq.~(\ref{eq16}) do satisfy the relation (\ref{eq21}).

Although the expression for the flux function obtained above is valid only at $z\to\infty$, the expression for the current function is valid all over the magnetic field lines close enough to the axis. With this function in hand, from Eq.~(\ref{eq12}) one can obtain the flux function at $z\to 0$, i.e. directly at the top of the polar gap,
\begin{equation}
f=\frac{\rho^2}{z^3}-\frac{3}{2}\frac{\rho^4}{z^5}+\frac{15}{8}\frac{\rho^6}{z^7}-\frac{203}{96}\frac{\rho^8}{z^9}+\dots, 
\label{eq22}
\end{equation}
which differs from the dipole expansion as
\begin{equation}
f-f_\mathrm{dip}\sim\frac{7}{96}\frac{\rho^8}{z^9}. 
\label{eq23}
\end{equation}
By means of Eq.~(\ref{eq22}), (the polar region of) the transverse current sheet at the neutron star surface is included into the global structure of the pulsar force-free magnetosphere.

\section{Discussion} 
\protect\label{s4}

Our paper presents one of the scarce attempts of analytic treatment of the pulsar equation. An idea to start such a consideration is motivated by the actual necessity to complement the recent multitudinous numerical results on the pulsar force-free magnetosphere with the physically grounded theoretical advances. In our previous paper \citep{p12}, we have suggested an empirical model of the pulsar axisymmetric force-free magnetosphere, which includes the polar, outer and slot gaps. The present paper starts the analytic description of this model and concentrates on the region close to the magnetic axis.

Since along the axis the current function is zero, in the closest vicinity of the axis it can be linearized. Consequently, in the axial region, the pulsar equation is essentially simplified and can be solved by separating variables. As a result, we have obtained the axial flux function (\ref{eq9}) valid at any altitude above the neutron star. It links the magnetic field below the stellar surface to that at the top of the polar gap and that at infinity. It is shown that at infinity the axial field of a force-free dipole asymptotically coincides with the field of a monopole, $f_\mathrm{mon}=2(1-z/\sqrt{z^2+\rho^2})\approx\rho^2/z^2$.

The axial flux function (\ref{eq9}) is used as a starting approximation in the treatment of the complete non-linear pulsar equation in terms of series in $\rho^2/z^2$. Taking into account the quasi-monopolar character of the magnetic flux at large radial distances yields unique asymptotic series for the flux and current functions in the polar region. At $z\to\infty$ both functions appear to differ slightly from those for the force-free monopole, $f-f_\mathrm{mon}\sim -\rho^6/8z^6$, $AA^\prime-AA^\prime_\mathrm{mon}\sim -4f^3$. At the top of the polar gap, the flux function slightly differs from that of a pure dipole, $f-f_\mathrm{dip}\sim 7\rho^8/96z^9$. Although the asymptotic series obtained do not seem to be suitable for direct quantitative applications, our results have several implications.

With the axial flux function in the form (\ref{eq9}), we have the second derivative in $\rho$ at the axis, $(\partial^2f/\partial\rho^2)_{\rho=0}=2\Phi(z)$, which can be regarded as an additional condition at a segment of the boundary in the elliptic-type problem associated with the pulsar equation. Within the framework of the quasi-reversal method of solving differential equations in partial derivatives, this may be used to compensate for the unknown part of the boundary or the unknown boundary condition. This is especially actual, since the full set of boundary conditions in the pulsar force-free magnetosphere with the plasma-producing gaps is still unknown.

The flux function at the top of the polar gap (\ref{eq22}) demonstrates that the magnetic field differs from the original dipolar one. Our result implies that it is the transverse current closing the pulsar circuit that is responsible for the magnetic field distortion. Then an exact form of the transverse current is sufficient to successively find the magnetic field at the top of the polar gap, the force-free current distribution across the polar-gap-controlled field lines and the force-free field up to infinity. Thus, the long-standing problem of the polar gap adjustment to the global magnetospheric structure is reduced to the problem of consistent incorporating the pulsar current circuit closure.

Generally speaking, the global structure of the force-free fields and currents is a basis for examining the microphysics of particle flows all over the magnetosphere. Of course, such a detailed study is beyond the framework of the present paper, but with the flux and current functions obtained we can find the net charge and poloidal current densities which supply the global magnetospheric configuration in the regions of interest. Retrieving the dimensional coefficients, we have
\begin{equation}
f=\frac{B_0R_\mathrm{NS}^3}{2R_\mathrm{L}^3}R^2\Phi(Z/R_\mathrm{L}), \quad AA^\prime =\frac{4f}{R_\mathrm{L}^2}, 
\label{eq24}
\end{equation}
where $B_0$ is the magnetic field strength at the neutron star surface, $R_\mathrm{NS}$ the stellar radius, $R_\mathrm{L}=c/\Omega$ the light cylinder radius, $\Omega$ the angular velocity of pulsar rotation, $R=\rho R_\mathrm{L}$, $Z=zR_\mathrm{L}$. Then for the charge and current densities,
\[
\rho_e=-\frac{\Omega}{4\pi c}\left(\frac{\partial^2f}{\partial R^2}+\frac{1}{R}\frac{\partial f}{\partial R}+\frac{\partial^2 f}{\partial Z^2}\right), 
\]
\begin{equation}
J_p=-\frac{cA^\prime}{4\pi R}\sqrt{\left(\frac{\partial f}{\partial R}\right)^2+\left(\frac{\partial f}{\partial Z}\right)^2}, 
\label{eq25}
\end{equation}
we have
\[
\rho_e=-\frac{\Omega B_0}{2\pi c}\frac{R_\mathrm{NS}^3}{R_\mathrm{L}^3}\Phi(Z/R_\mathrm{L}),
\]
\begin{equation} J_p=-\frac{\Omega B_0}{2\pi c}\frac{R_\mathrm{NS}^3}{R_\mathrm{L}^3}\Phi(Z/R_\mathrm{L})c,
\label{eq26}
\end{equation}
where $-\Omega B_0/2\pi c$ corresponds to the Goldreich-Julian charge density at the top of the polar gap, the counteraligned rotator is chosen, it is taken into account that $A^\prime=-2/R_\mathrm{L}$ and only the eldest terms in $R$ are kept. As can be seen from Eq.~(\ref{eq26}), $J_p=\rho_ec$.

The same result, $J_p/\rho_ec=1$, is also obtained with the asymptotic flux function at infinity (\ref{eq16}). Note that, despite the distinction from the field of a monopole, the equality is fulfilled exactly, just as in the monopolar case. For the flux function at the top of the polar gap (\ref{eq22}) we have $J_p/\rho_e c\approx 1-\rho^2/2z^3$. As could be expected, this coincides with the relation (A2) in \citet{t06} for the purely dipolar field, since the difference of the flux functions is very small (cf. Eq.~(\ref{eq23})). 

Note that in the polar region considered in the present paper the poloidal current density differs only slightly from the Goldreich-Julian current density, $0<1-J_p/\rho_ec\ll 1$. This particular result is in line with the stationary pair cascade inside the polar gap and is likely to testify against the backward particle flow along the axis. Although the backward particle flow is thought of as a necessary attribute of the pulsar magnetosphere \citep{l09}, it rather streams along the field lines controlled by the outer gap. The recent numerical simulations of the coexisting gaps in the pulsar magnetosphere show the backward particle flow along the axis \citep[see Fig.1 in][]{s12}, but the authors themselves admit that this, together with the non-zero accelerating electric field at the axis, is rather an artifact of their numerical method.

Our consideration have demonstrated that the asymptotic representations of the flux and current functions close to the magnetic axis are unique. In the polar region, these functions appear close enough to those obtained by means of numerical simulations in \citet{ckf99} and a number of subsequent works. This could be expected, since we have assumed the same boundary condition at the axis. Note that at larger polar angles the presence of the outer and slot gaps should markedly change the force-free field as compared to that known from the previous numerical simulations of the pulsar force-free magnetosphere. This will be a subject of our future study.

\section{Conclusions} 
\protect\label{s5}

We have considered the stationary axisymmetric force-free magnetosphere of a pulsar and performed the analytic treatment of the pulsar equation close to the magnetic axis. Linearization of the current function in the axial region leads to the self-consistent flux function with the approximately dipolar and monopolar behaviour at low and high altitudes, respectively. The axial flux function has been used to construct the solution of the non-linear pulsar equation at small polar angles in terms of series. Keeping in mind the quasi-monopolar character of the flux function at infinity, we have obtained unique asymptotic series for the flux and current functions, which differ from the monopolar ones as $f-f_\mathrm{mon}\sim -\rho^6/8z^6$, $AA^\prime-AA^\prime_\mathrm{mon}\sim -4f^3$. With the current function obtained, the flux function at the top of the polar gap appears to differ from the dipolar one as $f-f_\mathrm{dip}\sim 7\rho^8/96z^9$. This is attributed to the action of the transverse current closing the current circuit. Thus, our result includes the polar part of the transverse current sheet at the neutron star surface into the global magnetospheric structure.

Possible implications of our results can be summarized as follows.
\begin{enumerate}
 \item The axial flux function found implies an additional boundary condition at the axis, $(\partial^2f/\partial\rho^2)_{\rho=0}=2\Phi(z)$. This may, at least partially, compensate for the unknown boundary segments and boundary conditions in the pulsar force-free problem.
 \item The self-consistent magnetic field at the top of the polar gap may constrain the physics of current closure in pulsars and give new insights into the polar gap theory.
 \item Our results testify against the backward particle flow at small polar angles.
 \item The relation of the resultant net charge and poloidal current densities does not exclude the stationary cascade scenario in the polar region of the polar gap.
\end{enumerate}

The results of the present paper are the first step in the analytic description of our model of the pulsar force-free magnetosphere \citep{p12}, which consistently includes the plasma-producing gaps. As our model assumes the same axial boundary conditions as the classical one, in the polar region considered here our results are roughly in line with the numerical results obtained within the framework of the classical model. However, in other parts of the magnetosphere our model is essentially distinct, and the analytic description of fields and currents in these regions is of particular interest. This will be a subject of the subsequent papers.

\section*{Acknowledgements}
I am grateful to the anonymous referee for the interesting question and enlightening discussion.
The work is partially supported by the grant of the President of Ukraine (the project of the State Fund for Fundamental Research No. F35/554-2011).

\begin{appendix}
\section{Continuous transition across the polar gap \label{sa}}

We are going to find the solution of Eq.~(\ref{eq6}) at $z\gg z_0$, taking into account that at $z<z_0$ $u=\mathrm{exp}(-\lambda z)$. It is convenient to introduce a smoothed step function in the form
\begin{equation}
\chi(z-z_0)=\frac{(z-z_0)/h}{1+(z-z_0)/h},\quad h\ll 1.
\label{a1}
\end{equation}
Note that the results of this Appendix are not affected by the concrete way of smoothing the step function. The form (\ref{a1}) is chosen for the sake of simplicity and it is not expected to have anything in common with the true distribution of current inside the transverse current sheet.
With Eq.~(\ref{a1}) we reduce Eq.~(\ref{eq6}) to
\begin{equation}
(1+y)\frac{\mathrm{d}^2u}{\mathrm{d}y^2}-h^2\left[\lambda^2+\left(\lambda^2-\alpha\right)y\right]u=0,
\label{a2}
\end{equation}
where $y\equiv (z-z_0)/h$. The solution of Eq.~(\ref{a2}) reads
\[
u=2h\sqrt{\lambda^2-\alpha}(y+1)\mathrm{e}^{-h\sqrt{\lambda^2-\alpha}}
\]
\[\times
\left\{C_1M\left[1+\frac{2h}{\sqrt{\lambda^2-\alpha}},2,2h\sqrt{\lambda^2-\alpha}(y+1)\right]\right.\]
\begin{equation}
+C_2U\left.\left[1+\frac{2h}{\sqrt{\lambda^2-\alpha}},2,2h\sqrt{\lambda^2-\alpha}(y+1)\right]\right\},
\label{a3}
\end{equation}
where $M(a,b,x)$ and $U(a,b,x)$ are the degenerate hypergeometric functions. At $h\to 0$ one can make use of the relations
\[
M(\nu+1/2,2\nu+1,2x)=\Gamma(1+\nu)\mathrm{e}^x\left (\frac{x}{2}\right )^{-\nu}I_\nu(x),
\]
\begin{equation}
U(\nu+1/2,2\nu+1,2x)=\frac{\mathrm{e}^x}{\sqrt{\pi}}\left (2x\right )^{-\nu}K_\nu(x),
\label{a4}
\end{equation}
where $I_\nu(x)$ and $K_\nu(x)$ are the modified Bessel functions. Then, taking into account that
\begin{equation}
I_{1/2}(x)=\sqrt{\frac{2}{\pi x}}\mathrm{sh} x,\quad K_{1/2}(x)=\sqrt{\frac{\pi}{2x}}\mathrm{e}^{-x},
\label{a5}
\end{equation}
and returning to the variable $z$, we obtain finally
\begin{equation}
u=\mathrm{e}^{-\lambda z_0-(z-z_0)\sqrt{\lambda^2-\alpha}}.
\label{a6}
\end{equation}
The latter equation describes the function $u$ above the polar gap, $z>z_0$, which results from continuous passage through the infinitesimal transverse current sheet according to Eq.~(\ref{eq6}). At $z=z_0$ we have $u=\mathrm{e}^{-\lambda z_0}$, and at $z\gg z_0$ Eq.~(\ref{a6}) takes the form
\begin{equation}
u=\mathrm{e}^{-z\sqrt{\lambda^2-\alpha}}.
\label{a7}
\end{equation}

\end{appendix}

\clearpage

\clearpage

\end{document}